\documentclass[
  twocolumn,
  10pt,
  prd,
  nofootinbib,
  preprintnumbers,
  balancelastpage,
  superscriptaddress
]{revtex4-2}
\usepackage[utf8]{inputenc}
\usepackage{setspace,flushend}

\usepackage[dvips]{graphicx}
\usepackage[dvipsnames]{xcolor}
\definecolor{CiteBlue}{RGB}{45,52,151}
\usepackage[
    colorlinks=true,
    linkcolor=CiteBlue,
    urlcolor=CiteBlue,
    citecolor=CiteBlue
]{hyperref}

\usepackage[version=4]{mhchem}
\usepackage[capitalise]{cleveref}

\usepackage{bm}
\newcommand{\bb}[1]{\bm{\mathrm{#1}}}

\usepackage{mhchem}

\usepackage{siunitx}
\DeclareSIUnit{\year}{yr}
\DeclareSIUnit{\rpm}{rpm}
\newcommand{\eV}{\electronvolt}

\newcommand{\refcite}[1]{Ref.~\cite{#1}}
\newcommand{\refscite}[1]{Refs.~\cite{#1}}
\newcommand{\du}{\mathrm d}
\renewcommand{\Im}{\operatorname{Im}}
\renewcommand{\Re}{\operatorname{Re}}
\newcommand{\fermi}{\mathrm{F}}
\newcommand{\plasma}{\mathrm{p}}
\newcommand{\dm}{\chi}
\newcommand{\med}{\phi}
\newcommand{\el}{\mathrm{e}}
\newcommand{\mstar}{m_\el}
\newcommand{\loss}{\mathcal W}
\newcommand{\thinloss}{\mathcal V}
\newcommand{\erf}{\operatorname{erf}}

\begin{document}

\title{New constraints on dark matter from superconducting nanowires}

\author{Yonit Hochberg}
\email{yonit.hochberg@mail.huji.ac.il}
\affiliation{Racah Institute of Physics, Hebrew University of Jerusalem, Jerusalem 91904, Israel}

\author{Benjamin V. Lehmann}
\email{benvlehmann@gmail.com}
\affiliation{Center for Theoretical Physics, Massachusetts Institute of Technology, Cambridge, MA, USA}

\author{Ilya Charaev}
\email{charaev@mit.edu}
\affiliation{Department of Electrical Engineering and Computer
Science, Massachusetts Institute of Technology, Cambridge, MA, USA}
\affiliation{University of Zurich, Zurich 8057, Switzerland}

\author{Jeff Chiles}
\email{jeffrey.chiles@nist.gov}
\affiliation{National Institute of Standards and Technology, Boulder, CO, USA}

\author{Marco Colangelo}
\email{colang@mit.edu}
\affiliation{Department of Electrical Engineering and Computer
Science, Massachusetts Institute of Technology, Cambridge, MA, USA}

\author{Sae Woo Nam}
\email{nams@boulder.nist.gov}
\affiliation{National Institute of Standards and Technology, Boulder, CO, USA}

\author{Karl K. Berggren}
\email{berggren@mit.edu}
\affiliation{Department of Electrical Engineering and Computer Science, Massachusetts Institute of Technology, Cambridge, MA, USA}

\date{\today}
\begin{abstract}
Superconducting nanowires, a mature technology originally developed for quantum sensing, can be used as a target and sensor with which to search for dark matter interactions with electrons. Here we report on a 180-hour measurement of a tungsten silicide superconducting nanowire device with a mass of 4.3 nanograms. We use this to place new constraints on dark matter--electron interactions, including the strongest terrestrial constraints to date on sub-MeV (sub-eV) dark matter that interacts with electrons via scattering~(absorption) processes.
\end{abstract}

\maketitle

\section{Introduction}
The identity of the dark matter (DM) in the Universe remains one of the biggest mysteries of modern physics. After decades of theoretical and experimental focus on DM at the electroweak scale, attention has recently shifted to lighter masses, with sub-GeV DM capturing the limelight from both the theoretical~\cite{Essig:2011nj,Graham:2012su,Essig:2015cda,Lee:2015qva,Hochberg:2015pha,Hochberg:2015fth,Alexander:2016aln,Derenzo:2016fse,Hochberg:2016ntt,Kavanagh:2016pyr,Emken:2017erx,Emken:2017qmp,Battaglieri:2017aum,Essig:2017kqs,Cavoto:2017otc,Hochberg:2017wce,Essig:2018tss,Emken:2018run,Ema:2018bih,Geilhufe:2018gry,Baxter:2019pnz,Essig:2019xkx,Emken:2019tni,Hochberg:2019cyy,Trickle:2019nya,Griffin:2019mvc,Coskuner:2019odd,Geilhufe:2019ndy,Catena:2019gfa,Blanco:2019lrf,Kurinsky:2019pgb,Kurinsky:2020dpb,Griffin:2020lgd,Radick:2020qip,Gelmini:2020xir,Trickle:2020oki,Du:2020ldo} and experimental~\cite{Essig:2012yx,Tiffenberg:2017aac,Romani:2017iwi,Crisler:2018gci,Agnese:2018col,Agnes:2018oej,Settimo:2018qcm,Akerib:2018hck,Abramoff:2019dfb,Aguilar-Arevalo:2019wdi,Aprile:2019xxb,Barak:2020fql,Arnaud:2020svb,Amaral:2020ryn} perspectives. Direct detection of sub-GeV DM requires detectors with much lower thresholds than traditional experiments, and this has motivated the development of many novel detection techniques. Among the proposed detectors, superconductors~\cite{Hochberg:2015fth, Hochberg:2015pha,Hochberg:2019cyy} stand out: due to their exceptionally small band gaps of $\mathcal O(\SI{}{\milli\eV})$ and correspondingly small detection thresholds, these materials are capable of detecting light sub-MeV DM. In principle, they are sensitive to the scattering (absorption) of DM with mass as light as $\sim$\SI{1}{\kilo\eV} ($\sim$\SI{1}{\milli\eV}).

Realizing the full potential of superconducting detectors for DM will require additional technological developments \cite{doi:10.1063/5.0045990}. However, existing devices being used for other applications can already play a meaningful role for dark matter detection. Superconducting nanowire single-photon detectors (SNSPDs) are one such established sensor technology, with numerous applications from quantum sensing to telecommunications (see \textit{e.g.} \refscite{Grein2014,Grein2014b,Natarajan2009}). These devices are sensitive to the deposit of extremely small amounts of energy, with proven sub-eV thresholds and low dark count rates \cite{Hochberg:2019cyy,Verma:2020gso,doi:10.1063/5.0048049,Moroz,chen2020midinfrared,Marsili2012,chang2021midinfrared,VermaS,Wollman2017} and potential to measure the spectrum of energy deposits~\cite{Kong:2021}. Under certain conditions, they may even be sensitive to the direction of the deposited momentum \cite{Hochberg:2021ymx}. In \refcite{Hochberg:2019cyy}, we proposed to apply this mature technology for the first time to the DM hunt by using the SNSPDs simultaneously as the target and for readout: \textit{i.e.}, the SNSPD is both the material with which DM interacts and the sensor that registers the deposited energy and momentum.

\begin{figure}
    \includegraphics[width=.48\textwidth]{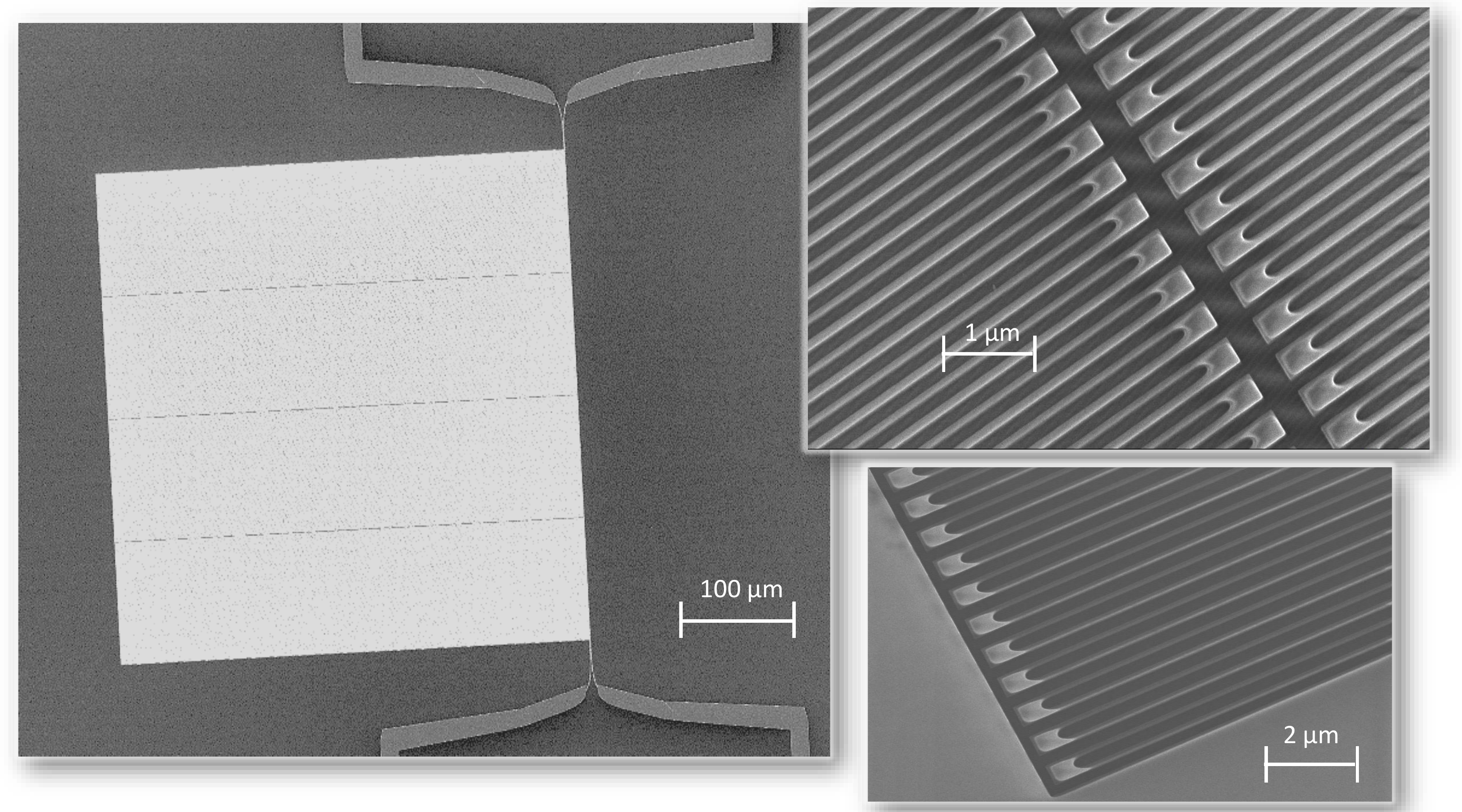}
    \caption{SEM images of the prototype WSi SNSPD device taken at different magnifications.
    \textit{Left:} the entire device with two contact pads and active area of \SI{400}{\micro\meter} by \SI{400}{\micro\meter}. \textit{Top right:} View of the detector area in the center. \textit{Bottom right:} Several individual nanowires.
    }
    \label{fig:exp}
\end{figure}

In this work, we report on a 180-hour measurement performed with a prototype SNSPD device that we use to place new bounds on DM, including the strongest terrestrial constraints to date on dark matter with sub-MeV (or sub-eV) masses that scatters with (or is absorbed by) electrons. For the first time, we evaluate bounds using a novel theoretical framework that accounts for the many-body physics of the detector and includes an enhancement due to the thin-layer geometry. 
Our results represent novel constraints on DM interactions from a superconducting detector system, realizing prospects envisioned nearly a decade ago and providing a new driver for the development of quantum sensing technology. We present a roadmap for the development of future experiments and demonstrate the prospects for SNSPDs to lead exploration of the light DM parameter space. Throughout this work we use natural units, where $c=\hbar=1$.

\section{Experimental setup}
\begin{figure}
    \includegraphics[width=.42\textwidth]{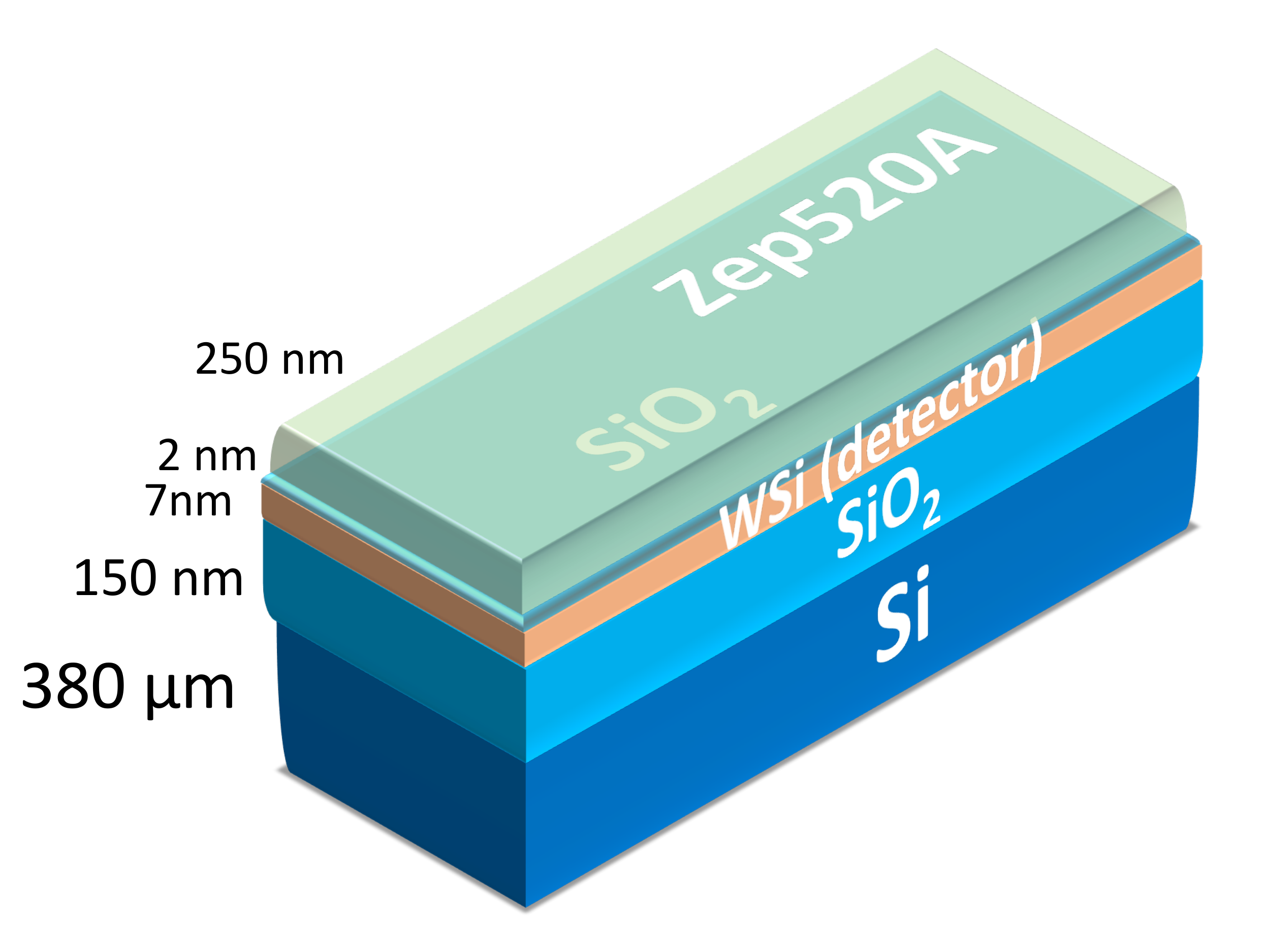}
    \caption{Schematic cross section of a single nanowire. Layers are not drawn to scale.}
    \label{fig:snspd-cross-section}
\end{figure}

SNSPDs operate by maintaining a bias current in a superconducting nanowire, keeping the device in the superconducting phase very near the edge of the superconducting transition. Under these conditions, any deposited energy above threshold can cause a portion of the device to undergo a transition to the normal metal phase, locally increasing the resistance of the wire. This results in a brief but significant voltage pulse that can be amplified and then read out. Typical events produce pulses with an amplitude of order \SI{1}{\milli\volt} lasting for several nanoseconds for absorbed energy ranging from \SI{0.1}{\milli\electronvolt} to \SI{10}{\electronvolt}. Further information on energy thresholds and calibration can be found in \refcite{bhargav2021metrology}.

Scanning electron microscope (SEM) images of our prototype device are shown in \cref{fig:exp}. The device is a square array of nanowires measuring \SI{400}{\micro\meter} on a side, with two contact pads for the readout electronics. Each nanowire in the array measures \SI{140}{\nano\meter} in width, and the spacing between each wire and the next is \SI{200}{\nano\meter}, corresponding to a pitch of \SI{340}{\nano\meter}. Each nanowire consists of several layers, illustrated in \cref{fig:snspd-cross-section}. The thin tungsten silicide (WSi) layer is the active detector layer, but the other layers still modify the detector response to deposited energy and momentum, as we discuss below.
The device was fabricated from a \SI{7}{\nano\meter}-thick WSi film which was sputtered on a \SI{150}{\nano\meter}-thick thermal silicon oxide film on a silicon substrate at room temperature with RF co-sputtering. Additionally, a thin \SI{2}{\nano\meter} Si layer was deposited on top of the WSi film in-situ to prevent oxidation of the superconductor. A layer of ZEP520A, a high performance positive tone electron beam resist, was spin-coated onto the chip at \SI{5000}{\rpm}, which ensured a thickness of \SI{335}{\nano\meter}. The ZEP520A pattern was then transferred to the WSi by reactive ion etching in \ce{CF_4} at 50~W. The ZEP520A thickness is estimated to be \SI{250}{\nano\meter} after etching and is left on the top surface.
\begin{figure}
    \includegraphics[width=.49\textwidth]{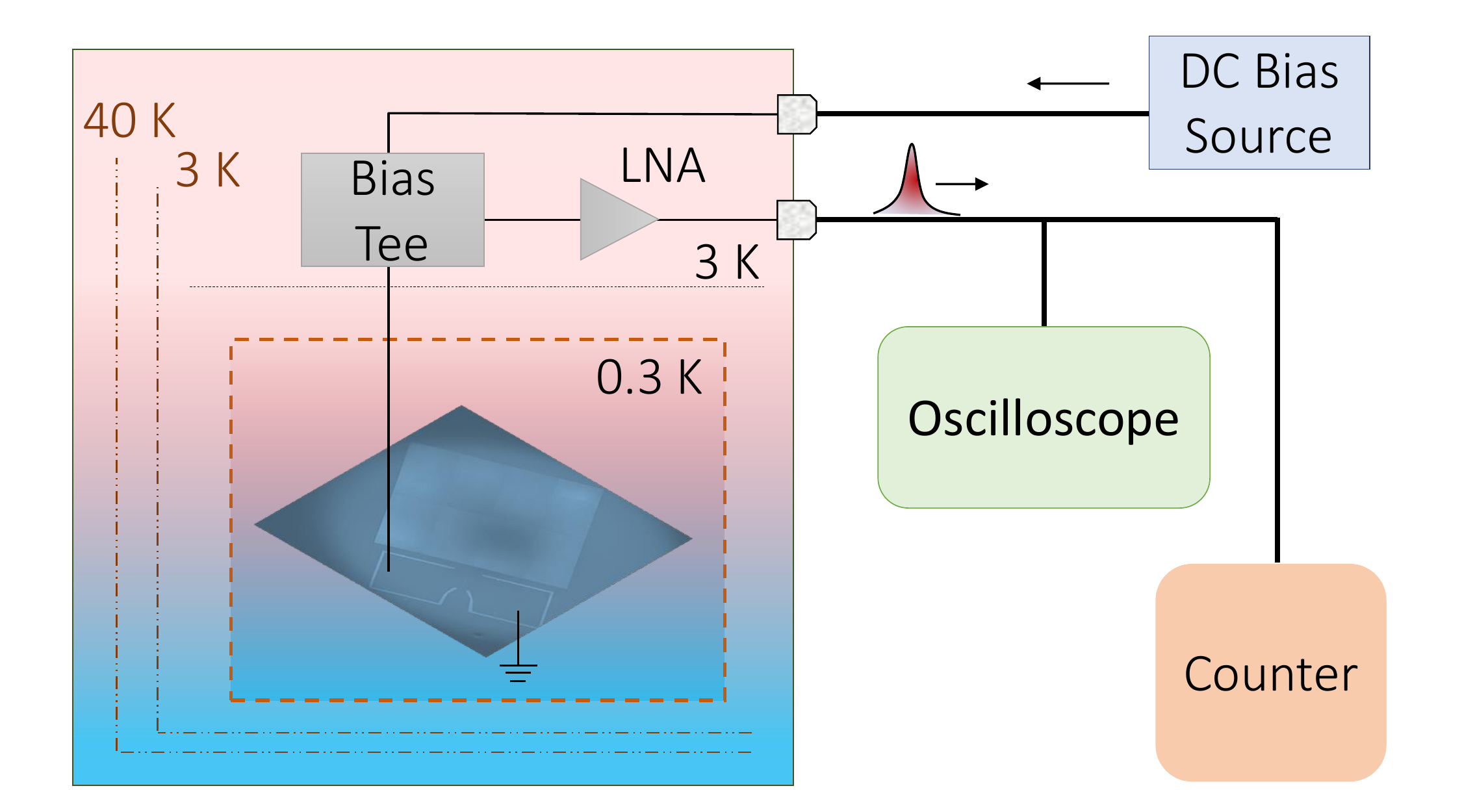}
    \caption{Sketch of the experimental setup. The prototype device was embedded in a light-tight box and cooled to a temperature of \SI{0.3}{\kelvin}. The high-frequency signal was carried out of the cryostat though a low-noise cryogenic amplifier to the read-out, while the DC path was connected to a low-noise voltage source. A low-temperature bias tee decoupled the high-frequency path from the DC bias path at the \SI{3}{\kelvin} stage.}
    \label{fig:set}
\end{figure}
The prototype device is contained inside a light-tight box at \SI{0.3}{\kelvin} as shown in \cref{fig:set}.  The signal was amplified at the \SI{3}{\kelvin} stage by cryogenic low-noise amplifiers with a total gain of \SI{56}{\deci\bel} and then sent to a pulse counter. To minimize the effect of blackbody illumination, the optical path was disconnected. The cryostat also has several layers of shielding at the \SI{3}{\kelvin} and \SI{40}{\kelvin} stages. For the science run, the bias current was fixed to \SI{4.5}{\micro\ampere}, and the device was exposed for 180 hours, with four dark counts observed. The device threshold is at most \SI{0.73}{\eV}. The observed dark counts may be due to cosmic ray muons, Cherenkov photons generated in the optical setup, or high-energy particles excited by radioactive decay events.
The data is further described in \refcite{Chiles:2021gxk}, which studies DM absorption in a haloscope configuration.

We use this data to set  world-leading bounds on DM--electron interactions, as explained below.

\begin{figure*}\centering
    \includegraphics[width=0.49\textwidth]{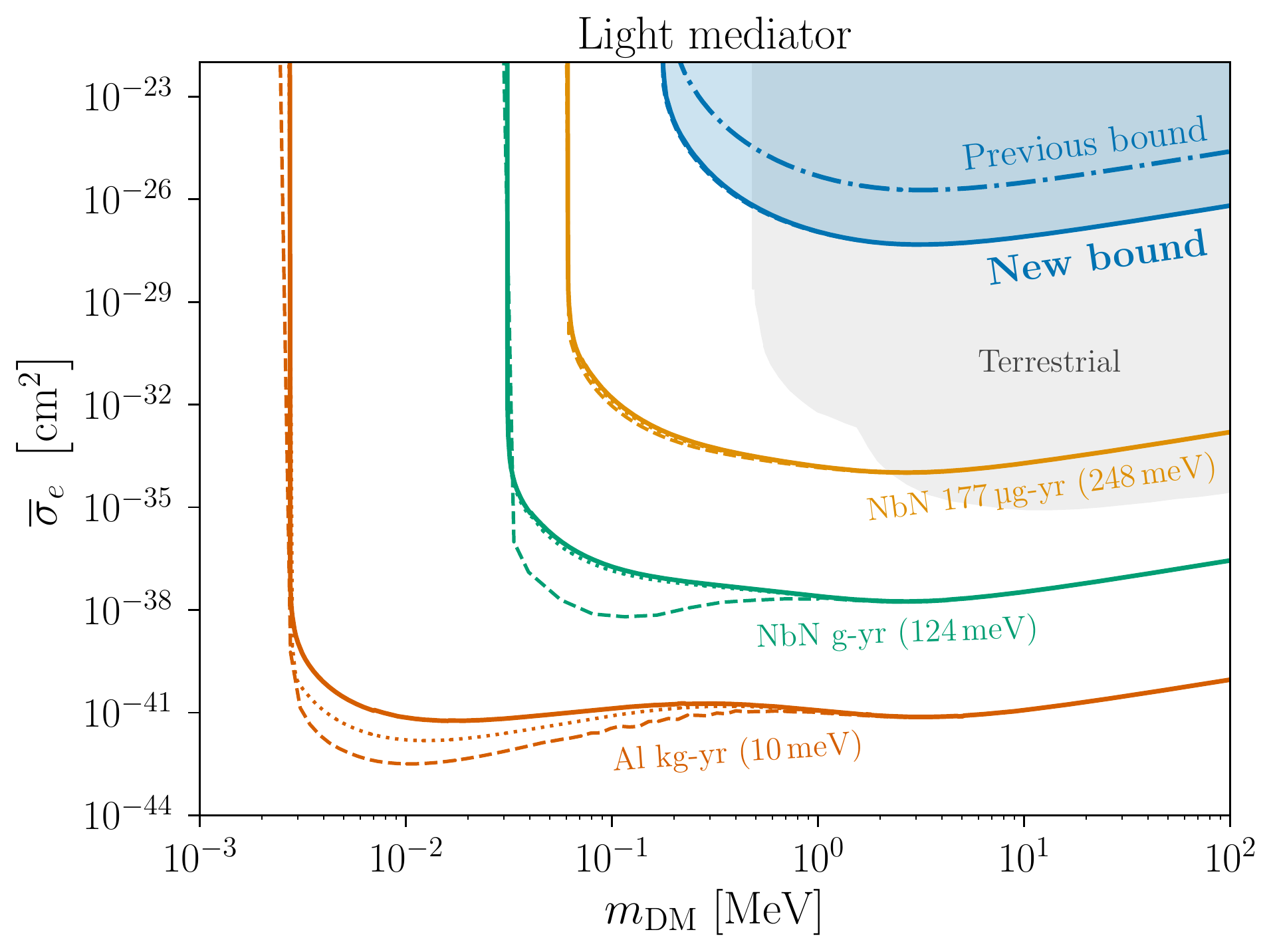} 
    \hfill
    \includegraphics[width=0.49\textwidth]{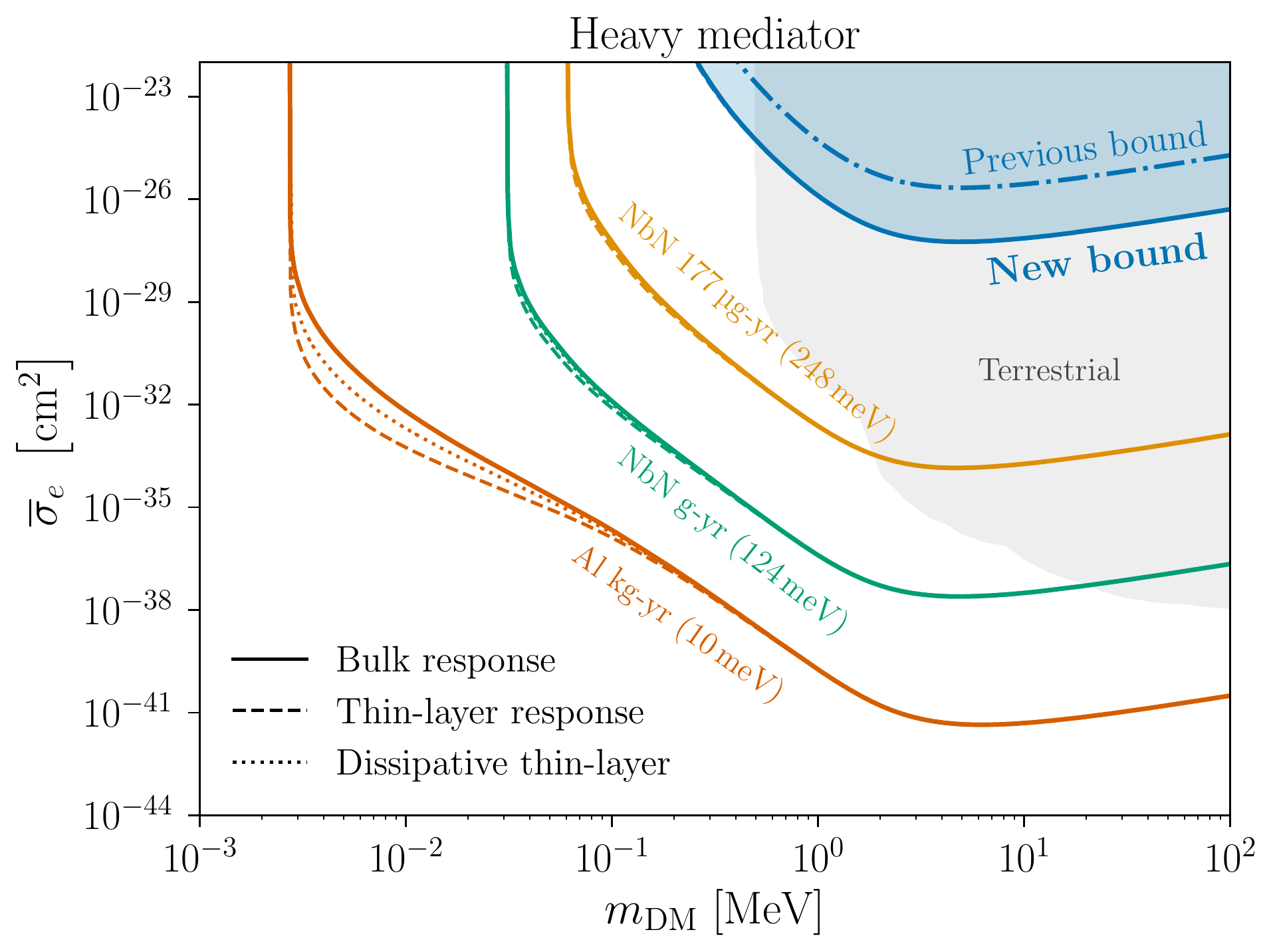} 
    \caption{
        New constraints and updated expected reach for DM--electron scattering in SNSPDs via light (\textit{left panel}) and heavy~(\textit{right panel}) mediators at 95\% C.L. as a function of DM mass.  The shaded blue region indicates the new bound placed by our prototype device with \SI{4.3}{\nano\gram} exposed for 180 hours with four dark counts observed. The dot-dashed blue curve indicates results from our previous run~\cite{Hochberg:2019cyy} with an exposure of \num{10000} seconds, now updated to include in-medium effects. Other curves show the projected reach for WSi, NbN, or Al targets with the indicated exposures and thresholds, assuming that sources of dark counts are eliminated. Solid curves conservatively neglect thin-layer enhancements. Dashed curves include these enhancements following \refcite{Lasenby:2021wsc}. Dotted curves conservatively include estimated effects of dissipation in neighboring layers (see text). The \SI{177}{\micro\gram} exposure corresponds to a $\SI{10}{\centi\meter}\times\SI{10}{\centi\meter}$  area of NbN at \SI{4}{\nano\meter} thickness and a 50\% fill factor, and 248~(124)~\SI{}{\milli\eV} threshold corresponds to a 5~(10)~\SI{}{\micro\meter} wavelength. In shaded gray we show the existing constraints from SENSEI~\cite{Barak:2020fql}, SuperCDMS HVeV~\cite{Amaral:2020ryn}, DAMIC~\cite{Aguilar-Arevalo:2019wdi}, Xenon10~\cite{Essig:2017kqs}, DarkSide-50~\cite{Agnes:2018oej}, and Xenon1T~\cite{Aprile:2019xxb}.
    }
    \label{fig:scat}
\end{figure*}

\section{DM interaction rate}
The concept of our experiment is that local DM particles may interact with the electrons in an SNSPD. In this case, a DM particle may occasionally exchange sufficient energy with these electrons to overcome the threshold of the detector, producing a count in the device when no other sources are present. In order to translate rate measurements of an SNSPD device to bounds on the DM--electron interactions, for both scattering and absorption processes, it is necessary to compute the rates of these processes in the detector.

For small energy and momentum transfers, electrons in the detector cannot be considered free particles, and the many-body physics of the target material becomes important. We compute the DM interaction rates using a new theoretical method recently developed by \refcite{Hochberg:2021pkt} (see also \refcite{Knapen:2021run}). This technique is based on the dielectric response of the target material and naturally incorporates the many-body physics of the detector, eliminating substantial uncertainties associated with first-principles approaches. The key input quantity, the \textit{dielectric function,} can be either measured experimentally or computed theoretically using established models from condensed matter physics.

\refscite{Hochberg:2021pkt,Knapen:2021run} determine the DM interaction rate assuming a bulk volume for the target. However, each unit of our prototype detector is composed of a stack of thin layers of different materials, as illustrated in \cref{fig:snspd-cross-section}. For a low-dimensional target system, or for heterogeneous systems with interfaces, the dielectric response of the detector is different from that of a bulk sample of material, and these differences should be accounted for in the rate. These effects are newly explored in \refcite{Lasenby:2021wsc}, which derives the DM interaction rate in a thin layer. In particular, if the layer width is small compared to the inverse momentum transfer in the interaction, the response of the layer itself is significantly modified, and features a new resonance for small energy deposits. Thus, the DM scattering rate per unit volume for a thin layer can be enhanced significantly with respect to a bulk detector.

Preliminary estimates suggest that the absorption rate is subject to even larger enhancements, but the approach of \refcite{Lasenby:2021wsc} cannot be directly applied in this kinematic regime, where the deposited momentum is much smaller than the deposited energy. We do not quantify this enhancement in this work, but leave this as a task for future experimental characterization.

The thin-layer interaction rate derived by \refcite{Lasenby:2021wsc} assumes that the detector layer is the only dissipative component of the system, such that energy deposited in any other layer is eventually dissipated there. However, experimental characterization of our prototype detector suggests that dissipation in the other layers is in fact significant: only large deposits far above the threshold in the other layers produce measurable events in the WSi layer. Thus, in what follows, we also show a conservative result that includes dissipation in all layers, and neglects deposits outside the detector layer. Further details are given in \cref{sec:appendix-geometry}. Our treatment yields a conservative bound on DM--electron interactions compared to what could be achieved with more complete knowledge of the prototype device response.  Future study of the prototype nanowire to accurately characterize sensitivity to energy deposits outside the WSi layer, as a function of their magnitude and location, will allow for even stronger DM limits.

We consider both DM scattering and absorption processes. For DM scattering, we place limits on the DM--electron scattering cross section. These hold for any spin-independent interaction that couples the DM to the electron density~\cite{Hochberg:2021pkt}, including both scalar and vector mediators. For DM absorption, we consider a relic dark photon and place limits on the kinetic mixing parameter $\kappa$. (See \cref{sec:appendix-interaction} for model details.)

\begin{figure}\centering
    \includegraphics[width=0.52\textwidth]{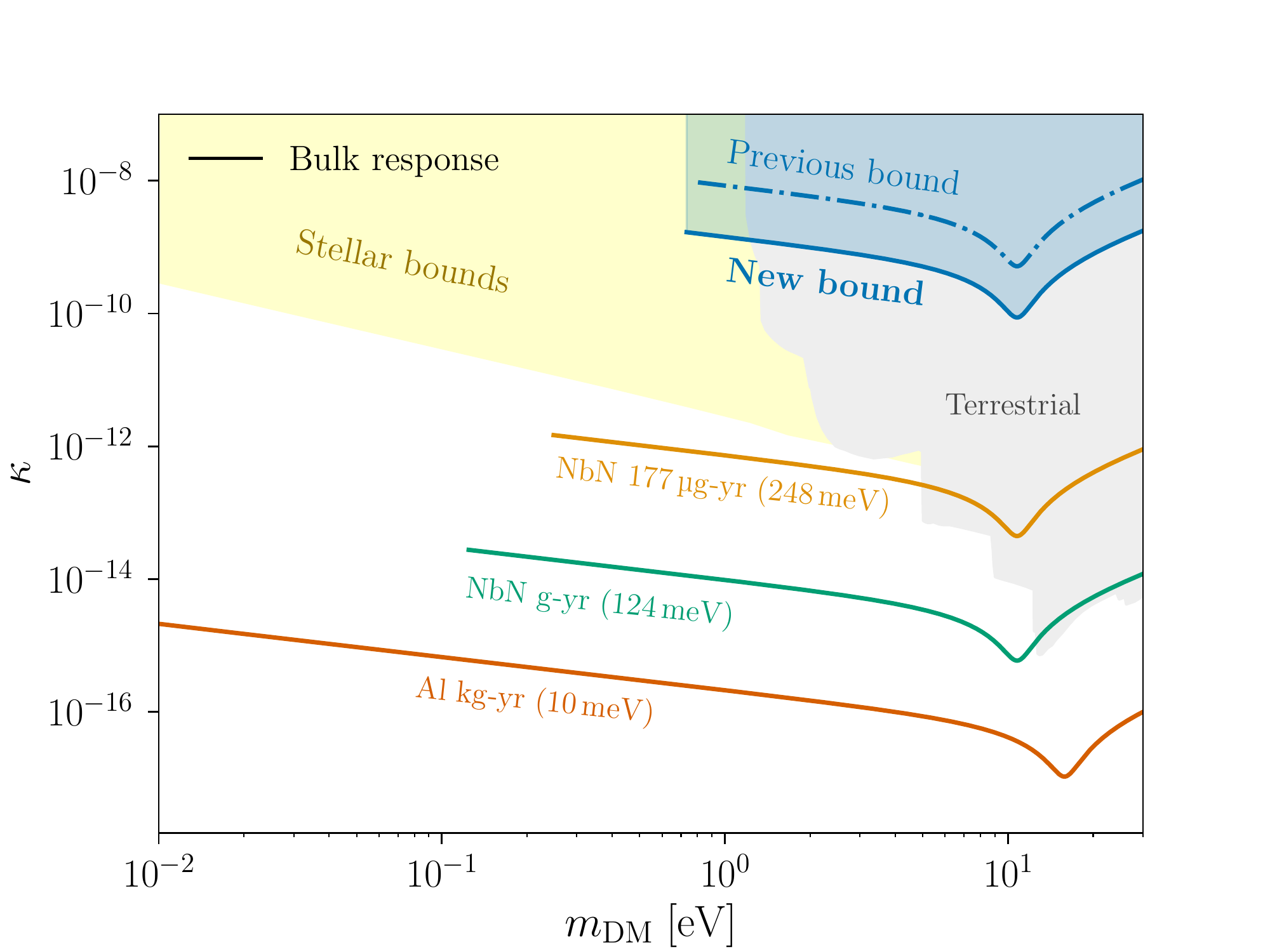}    
    \caption{New constraints and updated expected reach for DM absorption in SNSPDs as a function of DM mass, for a relic kinetically mixed dark photon. As in \cref{fig:scat}, the shaded blue region indicates the new bound at 95\% C.L., and other solid curves indicate projections for future experiments, neglecting possible geometric effects. The shaded gray region shows existing terrestrial constraints from Xenon data~\cite{An:2014twa}, SuperCDMS~\cite{Agnese:2018col}, DAMIC~\cite{Aguilar-Arevalo:2019wdi}, EDELWEISS~\cite{Arnaud:2020svb}, FUNK~\cite{FUNKExperiment:2020ofv} and SENSEI~\cite{Barak:2020fql}, while the yellow region indicates model-dependent stellar bounds~\cite{An:2013yua,An:2014twa,An:2020bxd}.}
    \label{fig:absorption}
\end{figure}

\section{Results}
Our new constraints are summarized in \cref{fig:scat} for DM--electron scattering with light and heavy mediators (left and right panels, respectively), and in \cref{fig:absorption} for DM absorption. Existing terrestrial constraints are shown in shaded gray, and model-dependent stellar constraints are shown in yellow. (Other model-dependent cosmological constraints may also apply; see \textit{e.g.} \refscite{Nguyen:2021cnb,Buen-Abad:2021mvc,Giovanetti:2021izc}.) Our previous nanowire bounds~\cite{Hochberg:2019cyy}, updated to incorporate in-medium effects via the dielectric formalism, are indicated by dot-dashed blue curves. Notably, our prototype detector already provides the strongest constraints to date on the electronic interactions of sub-MeV (sub-eV) DM via scattering (absorption) processes, with an exposure of only $\SI{4.3}{\nano\gram}\times\SI{180}{\hour}$ or equivalently \SI{8.8e-14}{\kilo\gram.\year}. We also show projections for future SNSPD experiments with larger exposures in NbN and Al detectors. All bounds and projections are given at 95\% confidence level (C.L.) for one-sided Poisson statistics and computed using the Lindhard model for the dielectric function \cite{dressel2002electrodynamics}, which agrees well with available measurements at zero momentum transfer.

Scattering results are shown in terms of a reference cross section $\bar\sigma_e = \frac1\pi\mu_{e\dm}^2g_e^2g_\dm^2\bigl[(\alpha_{\mathrm{EM}}m_e)^2 + m_\med^2\bigr]^{-2}$, where $\mu_{e\dm}$ is the reduced mass of the DM--electron system; $g_e$ and $g_\dm$ are the couplings of the mediator to the electron and DM, respectively; and $\alpha_{\mathrm{EM}} \approx 1/137$ is the fine structure constant. Absorption results are shown in terms of the size of the kinetic mixing $\kappa$ of a dark photon---essentially its coupling to the electromagnetic current. We take the Fermi energy $E_\fermi$ to be \SI{7}{\eV} in both WSi and NbN, and we take the densities to be \SI{9.3}{\gram/\centi\meter^3} and \SI{8.4}{\gram/\centi\meter^3}, respectively. The Fermi energy and density of Al are taken to be \SI{11.7}{\eV} and \SI{2.7}{\gram/\centi\meter^3}, respectively. We assume a local DM density of \SI{0.3}{\giga\eV/\centi\meter^3} with velocities distributed according to the Standard Halo Model, \textit{i.e.,} with probability density $f_\dm(\bb v) \propto \Theta(v_{\mathrm{esc}} - |\bb v|)\exp\left[-(\bb v+\bb v_{\mathrm{E}})^2/v_0^2\right]$. We take $v_0 = \SI{220}{\kilo\meter/\second}$, $v_{\mathrm{E}} = \SI{232}{\kilo\meter/\second}$, and $v_{\mathrm{esc}} = \SI{550}{\kilo\meter/\second}$.

The impressive reach for scattering and absorption at the smallest masses is due to the low device threshold of \SI{0.73}{\eV}, assisted by its low dark count rate. Future realizations of this experiment may be able to achieve substantially lower thresholds, sensitive to much lower masses. The projections for the reach of future NbN detectors assume thresholds of 248 and \SI{124}{\milli\eV}, which would extend the experimental reach to DM masses of order 50--\SI{100}{\kilo\eV}. Indeed, sensitivity at the \SI{10}{\micro\meter}-wavelength scale---corresponding to a \SI{124}{\milli\eV} threshold---has already been demonstrated in SNSPDs~\cite{Verma:2020gso}. We also show the projected reach for a superconducting Al detector with a \SI{10}{\milli\eV} threshold. Such a detector would be capable of detecting DM with mass of order $\sim$\SI{}{\kilo\eV}, below which structure formation considerations rule out fermionic DM~\cite{Tremaine:1979we,Boyarsky:2008ju,Boyarsky:2008xj}.

Solid curves are computed neglecting thin-layer effects, \textit{i.e.}, treating the detector as a bulk volume. Dashed and dotted curves show the projections including these effects: dashed curves neglect dissipation in the other layers, following \refcite{Lasenby:2021wsc}, while dotted lines incorporate this dissipation in the most conservative form. (See \cref{sec:appendix-geometry} for details.) Geometric effects do not significantly affect the reach of the constraints for the current experimental configuration, but these effects are an important consideration for future experimental design: thin-layer effects were not exploited in the original design of the prototype, and have arisen incidentally from the necessarily low-dimensional structure of SNSPDs. Sensitivity of the WSi detector layer to deposits in other layers of the device may allow for enhanced reach even at high DM masses, effectively increasing the detector volume. Such sensitivity may be possible for deposits far above threshold, and could be quantified experimentally. Deliberate optimization of the target geometry may enable even more significant enhancements, particularly in the absorption rate.

The geometric effects included in this work are estimated in a simplified framework. We do not quantify the geometric effects on the absorption rate here, and in the case of scattering, additional corrections may arise from the lower layers of the geometry in \cref{fig:snspd-cross-section} or from local-field corrections \cite{Adler:1962,Wiser:1963}. The accurate impact of the geometry of the device on the DM interaction rate can be quantified experimentally in a robust manner, and is expected to further improve the reach.

\section{Discussion}
We have reported on a new search for DM--electron scattering and absorption in a prototype SNSPD detector. Our results place the strongest terrestrial constraints to date on DM--electron interactions for sub-MeV (sub-eV) masses for scattering (absorption) processes. 
This is the first time that superconducting detectors have been used to probe unconstrained parameter space for DM scattering, a crucial milestone in the program of light DM searches that heralds significant collaboration between the DM and quantum-sensing communities.
The constraints presented in this work are computed using the dielectric function formalism, accounting for the many-body physics of the detector material, and we have also accounted for geometric effects that can significantly enhance the predicted DM interaction rate.

Our small-scale prototype is able to exceed previous experimental constraints thanks to the remarkably low \SI{0.73}{\eV} threshold of the SNSPD detector, along with its extremely low dark count rate. Future iterations of this experiment promise to reach even lower thresholds with even lower dark count rates. At present, we place constraints on DM interactions assuming that the dark counts are due to backgrounds. In the future, experimental improvements will allow the use of rate modulation~\cite{Freese:2012xd,Lee:2013xxa} and possibly even spectroscopic measurements~\cite{Kong:2021} to differentiate between backgrounds and a DM signal. The SNSPD platform is being heavily developed for numerous applications in quantum sensing and precision metrology, and given the rapid pace of development, \cref{fig:scat,fig:absorption} can be treated as a realistic indication of the reach of future experiments. The Al projections, with their \SI{10}{\milli\eV} thresholds, represent an ambitious target: achieving such thresholds will require considerable technological development, but there is no fundamental obstacle to constructing such a device.

An additional important challenge is to scale the prototype device to a large-scale experiment. Thus far, SNSPD devices are small: our nanogram-scale prototype is typical. Sensitivity to cross sections as small as those probed by experiments at higher DM masses will require significantly larger detectors at the gram scale and beyond. While the electron lithography techniques used to fabricate our prototype do not scale easily to larger devices, it is possible that optical lithography or other technologies would enable the production of a larger detector.

Finally, future experiments will be in a position to leverage geometric enhancements to the interaction rate. Our prototype detector was designed to demonstrate the capabilities of SNSPDs for DM detection with existing technology and fabrication techniques, and such geometric enhancements were not a design consideration. However, the theoretical methods introduced by \refscite{Hochberg:2021pkt,Knapen:2021run,Lasenby:2021wsc} make it possible to accurately compute these geometric effects when designing future detectors. The phenomenology of thin layers and interfaces has been studied thoroughly in the condensed matter literature, and this should allow for the fabrication of designer materials or heterostructures with highly customized dielectric responses. Such materials could feature even larger geometric enhancements to the DM interaction rate, allowing near-future experiments to delve deep into uncharted parameter space.

\begin{acknowledgments}
We thank Robert Lasenby for sharing preliminary results regarding geometric effects on the interaction rate and for helpful comments on a draft version of this manuscript. The work of Y.H. is supported by the Israel Science Foundation (grant No. 1112/17), by the Binational Science Foundation (grant No. 2016155), by the I-CORE Program of the Planning Budgeting Committee (grant No. 1937/12), and by the Azrieli Foundation. The work of B.V.L. is supported by DOE grant No. DE-SC0010107, by the Josephine de Karman Fellowship Trust, and by the MIT Pappalardo Fellowship. The experimental work on this effort at MIT was supported by the DOE under the QuantiSED program, grant No. DE-SC0019129. Work on data analysis and manuscript preparation at MIT was supported by the Fermi Research Alliance, LLC (FRA) and the US Department of Energy (DOE) under contract No. DE-AC02-07CH11359. The MIT co-authors thank Brenden Butters for technical assistance.
\end{acknowledgments}

\appendix

\section{DM interaction rate}
\label{sec:appendix-interaction}

We compute the rate of DM--electron scattering and absorption events using the recently-developed loss function formalism \cite{Hochberg:2021pkt,Knapen:2020aky} (sometimes called the dielectric formalism). This calculation is conceptually different from most experimental reach projections in the literature.

\subsection{Loss function formalism}

In the traditional approach, the DM scattering rate is computed from the microphysical scattering cross section between a DM particle and a single free electron. However, in the relevant range of energy and momentum transfers, electrons in detectors are generally not free. This is simultaneously an advantage and a difficulty for electron recoil experiments: on the one hand, DM scattering can induce transitions between electronic eigenstates whose kinematics are more favorable for detection. On the other hand, predicting the rate involves additional complications. In principle, to predict the DM scattering rate, one should compute transition rates between eigenstates of the material, but predicting the corresponding electronic wavefunctions is very challenging. In-medium effects can screen the DM--electron interaction, and other complicated many-body effects can modify the rate in either direction. Recent calculations account for these effects from first principles using sophisticated techniques such as density functional theory. However, the projected DM scattering rate is then subject to significant uncertainty associated with the modeling of the target material.

The recently-developed loss function formalism~\cite{Hochberg:2021pkt,Knapen:2020aky}, eliminates these uncertainties and provides a universal interpretation for in-medium effects across a wide variety of DM models. This approach is based on the fact that the response of the material to a deposited energy $\omega$ and momentum transfer $\bb q$ is independent of the nature of the interaction, as long as the interaction is weak and couples to the electron density. Under these conditions, the scattering rate can be written in terms of the non-relativistic interaction potential and a \textit{response function} that characterizes the physics of the material. The response function $\loss$ for scattering and absorption is known as the \textit{loss function,} and can be written in terms of the complex dielectric function $\epsilon$ as $\loss[\epsilon](\bb q,\omega) = \Im[-1/\epsilon(\bb q,\omega)]$. In terms of the loss function, the scattering rate becomes
\begin{equation}
    \label{eq:loss-function-rate}
    \Gamma = \int\frac{\du^3\bb q}{(2\pi)^3}|V(\bb q)|^2\left[
        2\frac{q^2}{e^2}
        \loss[\epsilon](\bb q,\,\omega)
    \right]
    \,,
\end{equation}
where $V(\bb q)$ is the DM--electron interaction potential, $e$ is the charge of the electron, and $\omega_{\bb q} = \bb q\cdot v_\dm - q^2/(2m_\dm)$, with $v_\dm$ and $m_\dm$ the DM velocity and mass, respectively.

The advantage of this approach is that the dielectric function has been studied extensively in the condensed matter literature as a key determinant of materials' optical properties \cite{dressel2002electrodynamics}. In particular, $\loss$ can be measured experimentally, removing all of the uncertainties associated with the material physics of the target system. Moreover, there are several established models that can approximate the dielectric function in different regimes of energy and momentum transfer. Thus, even in the absence of experimental data for a particular target material, it is possible to quickly and accurately predict the DM scattering rate including all in-medium effects.

In \cref{eq:loss-function-rate}, the physics of the target material is separated from the physics of the DM--electron interaction, and the latter enters only through the non-relativistic interaction potential $V(\bb q)$. For a spin-independent interaction, the scattering rate is independent of any other details of the interaction structure: the non-relativistic interaction potential in \cref{eq:loss-function-rate} takes the form
    $V(\bb q) = g_\dm g_e/(q^2 + m_\med^2)$,
where $g_\dm$ and $g_e$ are the couplings of the mediator to the DM and the electron, respectively; and $m_\med$ is the mediator mass, whether a scalar or a vector. The influence of the microphysical DM--electron interaction on the overall scattering rate is thus limited to the mediator mass and effective couplings. In particular, the response of the material does not depend on the nature of the interaction. With the interaction potential as above, the total event rate at fixed $\omega$ can be written in the form
\begin{equation}
    \mathcal R =
    \frac{2\rho_\dm g_\dm^2g_e^2}{e^2\rho_{\mathrm{D}}m_\dm}
    \int\frac{\du^3\bb q}{(2\pi)^3}
    \frac{1}{q^3}\int\du\omega\,f_\dm\left[v(q,\omega)\right]
    \,\loss[\epsilon](\bb q, \omega)
    \,,
\end{equation}
where $\rho_{\mathrm{D}}$ is the detector density, $\rho_\dm$ is the local DM density, $f_\dm$ is the DM velocity distribution, and $v(q,\omega) \equiv \omega/q + q/(2m_\dm)$.

This reorganization of the scattering rate calculation has clarified significant confusion in earlier literature regarding the dependence of in-medium effects on the nature of the DM--electron interaction. \refcite{Hochberg:2015fth} observed that in the case of a kinetically-mixed dark photon, the material response screens the interaction in exactly the same way that conductors screen applied electric fields. Later, \refcite{Gelmini:2020xir} pointed out that a similar effect can be derived in the case of a scalar-mediated interaction (see also \refcite{Mitridate:2021ctr}). The loss function formalism demonstrates immediately that the material response is identical for \textit{any} spin-independent interaction that couples to electron density, and thus the same calculation applies whether the mediator is a scalar or a vector.

Moreover, since any such interaction exhibits the same material response, the response function can be measured with electromagnetic interactions in the laboratory and then applied to predict scattering rates for DM--electron interactions. Although laboratory probes couple to the charge density rather than the electron density, energy losses in this regime are generally dominated by electronic interactions, and thus the experimentally-measured dielectric function is a very good approximation of the material response to DM scattering. The dielectric function, in turn, has been studied thoroughly in the condensed matter literature, and there are several established models that can approximate the dielectric function in different regimes of energy and momentum transfer. Thus, even in the absence of experimental data for a particular target material, it is possible to quickly and accurately predict the DM scattering rate including all in-medium effects.

Studying DM scattering with the dielectric function makes it possible to classify the different types of in-medium effects in the same language used by the condensed matter community. The screening of the interaction, for instance, is identical to screening in ordinary electromagnetism: a perturbation in the charge density induces subsequent fluctuations which partially cancel the applied potential. Less trivial phenomena are also naturally accommodated in this language. In particular, typical metals exhibit a resonance at non-zero $\omega$ and small $\bb q$ corresponding to the excitation of plasmons \cite{dressel2002electrodynamics}. Plasmons are eigenstates of the material that arise only as collective modes in the charge density. Thus, although the plasmon resonance can dramatically enhance the scattering rate, it is invisible in the single-particle formalism. The dielectric formalism includes plasmons automatically: the standard analytical approximations to the dielectric function account for such resonances, and, of course, experimental measurements naturally include \textit{all} collective modes that contribute to the scattering rate.

The loss function is readily measured by X-ray or electron scattering in the relevant regime of energy and momentum transfers. However, to our knowledge, no data is yet available for the loss function in WSi at the relevant values of $\bb q$ and $\omega$. Therefore, in this work, we compute the loss function using the well established Lindhard model \cite{dressel2002electrodynamics}. In the Lindhard model, also known as the random phase approximation or the free electron gas model, the loss function can be written in closed form in the low-temperature limit as
\begin{widetext}
\begin{equation}
    \label{eq:e-rpa}
    \epsilon_{\mathrm{L}}(\bb q,\omega) =
    1 + \frac{3\omega_\plasma^2}{q^2v_\fermi^2}\Biggl\{
        \frac12 + \frac{k_\fermi}{4q}\left(
            1 - Q_-^2
        \right)\operatorname{Log}\left(
            \frac{Q_- + 1}
                 {Q_- - 1}
        \right)
        +
        \frac{k_\fermi}{4q}\left(
            1 - Q_+^2
        \right)\operatorname{Log}\left(
            \frac{Q_+ + 1}
                 {Q_+ - 1}
        \right)
    \Biggr\}
    \,,
\end{equation}
\end{widetext}
where $\omega_\plasma = (4\pi\alpha n_\el/\mstar)^{1/2}$ is the plasma frequency, for $n_\el$ the number density of electrons; $k_\fermi$ is the Fermi momentum; $v_\fermi = k_\fermi / \mstar$ is the Fermi velocity; and $Q_\pm = q/(2k_\fermi) \pm \omega/(qv_\fermi)$. The Lindhard dielectric function exhibits a resonance at the plasma frequency $\omega_\plasma$. In the form above, this resonance is present but infinitely narrow. A non-zero width is obtained under the replacement $\omega \to \omega + i/\tau$, where the excitation lifetime $\tau$ can be fitted to experimental data. Such a width may enhance the loss function at deposits very far from the peak of the resonance \cite{Hochberg:2021pkt}. In this work, we estimate $1/\tau = \frac1{10}\omega_\plasma$, a typical width for a metal.

Each nanowire contains layers of Si and \ce{SiO2} in addition to WSi. While these layers do not enter into the scattering rate of \cref{eq:loss-function-rate}, they do play a role in the thin-layer effects discussed below. The dielectric function of Si can be approximated using the Lindhard model with $E_\fermi = \SI{18.9}{\eV}$, which originates from a phenomenological fit \cite{Hochberg:2021pkt}. For \ce{SiO2}, we use the fit provided by \refcite{Kischkat:12}. We model the ZEP520A top layer with a constant and real dielectric function, taking the (real) index of refraction to be $\sim$1.5.

The loss function formalism can also be used to predict absorption rates. For absorption, we consider a fiducial theory of a dark photon $A'_\mu$, with field strength $F'_{\mu\nu} \equiv \partial_\mu A'_\nu - \partial_\nu A'_\mu$, kinetically mixed with the Standard Model photon. That is, we assume a Lagrangian of the form
\begin{equation}
    \mathcal L \supset -\tfrac12\kappa F_{\mu\nu}F^{\prime\mu\nu}
    .
\end{equation}
The absorption rate per unit volume can then be written as 
\begin{equation}
    \label{eq:absorption-rate}
    \Gamma_{\mathrm{A}} = \kappa^2m_\dm\loss(\bb p_\dm, m_\dm)\,,
\end{equation} 
where $m_\dm$ is the DM mass and $\bb p_\dm$ is the momentum of the incoming DM particle. The kinetic mixing parameter $\kappa$ is the quantity that we bound in our experiment (see \cref{fig:absorption}).

We model the DM velocity distribution $f_\dm$ using the Standard Halo Model, with a distribution function of the form $f_\dm(\bb v) \propto \Theta(v_{\mathrm{esc}} - |\bb v|)\exp\left[-(\bb v+\bb v_{\mathrm{E}})^2 / v_0^2\right]$ and with parameter values $v_0 = \SI{220}{\kilo\meter/\second}$, $v_\mathrm{E} = \SI{232}{\kilo\meter/\second}$, and $v_{\mathrm{esc}} = \SI{550}{\kilo\meter/\second}$. However, to compute the bulk interaction rate on equal footing with the thin-layer interaction rate, we modify this approach slightly. The thin-layer rate depends not only on the DM speed, but on the direction of $\bb v_\dm$ with respect to the plane of the layer. Therefore, following \refcite{Lasenby:2021wsc}, we compute the interaction rate using the component of the DM velocity along a fixed axis, averaging over orientations with respect to the DM halo. This produces a DM speed distribution of the form
\begin{widetext}
\begin{equation}
    f_\dm(v) \propto \begin{cases}
        \frac{\pi^{1/2}}{4}\left[
            \erf\left(\frac{v_{\mathrm E} - v}{v_0}\right) + 
            \erf\left(\frac{v_{\mathrm E} + v}{v_0}\right)
        \right] - \exp\left(-v_{\mathrm{esc}}/v_0^2\right)
        & v < v_{\mathrm{esc}} - v_{\mathrm E}
        \\[0.25cm]
        \frac{\pi^{1/2}}{4}\left[
            \erf\left(\frac{v_{\mathrm{esc}}}{v_0}\right) + 
            \erf\left(\frac{v_{\mathrm E} - v}{v_0}\right)
        \right] - 
            \frac{v_{\mathrm E} + v_{\mathrm{esc}} - v}{2v_0}
            \exp\left(-v_{\mathrm{esc}}^2/v_0^2\right)
        & v_{\mathrm{esc}} - v_{\mathrm E} < v
            < v_{\mathrm{esc}} + v_{\mathrm E}
        \\[0.25cm]
        0 & v > v_{\mathrm{esc}} + v_{\mathrm E}.
    \end{cases}
\end{equation}
\end{widetext}
We use this distribution for both the bulk and thin-layer rate computations.

\section{Geometric enhancement to the interaction rate}
\label{sec:appendix-geometry}
The rate of \cref{eq:loss-function-rate} is written in a form appropriate for the scattering rate in a bulk volume. However, for thin layers, the dielectric response of the detector is different from that of a bulk sample of material. In particular, the relationship between the scattering rate and the dielectric function is modified: $\loss[\epsilon]$ is replaced by a new response function $\thinloss[\epsilon]$. This can significantly influence the DM interaction rate. This thin-layer response function can still be measured experimentally, but in the absence of experimental data, it is also possible to predict $\thinloss[\epsilon]$ given a model for the dielectric function $\epsilon$.

These effects are newly explored in \refcite{Lasenby:2021wsc}. \refcite{Lasenby:2021wsc} derives a function $R[\epsilon]$ such that $\thinloss = \frac1d\Re(R)$, where $d$ is the thickness of the detector layer (WSi in our prototype), and shows that the scattering rate per unit volume is exactly as given in \cref{eq:loss-function-rate} with the replacement $\loss \to \thinloss$. The response function $\thinloss$ is determined by solving the Poisson equation subject to the appropriate boundary conditions for a perturbing source with charge density $\rho = \rho_0e^{i(\bb q\cdot\bb x - \omega t)}$ and evaluating the time-averaged power deposited in each layer. Schematically, one makes the ansatz $\phi = \psi(z)e^{i(\bb q\cdot x - \omega t)}$, where $z$ is the coordinate normal to the layers. Then the Poisson equation reduces to an equation for $\psi(z)$, with the form
\begin{equation}
    -q^2\psi(z) + 2iq_z\psi'(z) + \psi''(z) = -\rho_0/\epsilon(z)
    .
\end{equation}
After imposing the appropriate boundary conditions and solving for $\psi$, the thin-layer loss function can be written as
\begin{equation}
    \label{eq:thin-loss}
    \thinloss = \frac{q^2}{d}\Re\left[
        -i\frac{1}{\rho}\int\du z\left(
            i\psi(z) + \frac{q_z}{q^2}\psi'(z)
        \right)
    \right]
    .
\end{equation}
Note that the integral in \cref{eq:thin-loss} is taken over all space, and the integrand has support outside the detector layer.

For a layer of thickness $d \ll q$, the resonance at the plasma frequency is suppressed compared to the bulk loss function. However, the thin-layer loss function exhibits a second resonance at smaller deposits, at $\omega \sim (qd/2)^{1/2}\omega_\plasma$, in the most important kinematic regime for light DM scattering. Thus, the DM scattering rate per unit volume for a thin layer can be enhanced significantly with respect to a bulk detector. Like the loss function $\loss$, the thin-layer response function $\thinloss$ is measurable for a particular target system.

One can make a first estimate of the geometric enhancements to absorption by assuming that the relationship between absorption and scattering is preserved, \textit{i.e.}, that the bulk response function $\mathcal W$ in \cref{eq:absorption-rate} can also be replaced with the thin-layer response function $\mathcal V$. An estimate carried out in this manner suggests that the absorption rate can be enhanced by one or two orders of magnitude in some regimes. However, \cref{eq:thin-loss} is derived under the assumption that the momentum transfer $q$ is much larger than the deposited energy $\omega$, which is not the case for absorption. Thus, we do not show thin-layer curves in \cref{fig:absorption}, and leave a quantitative treatment to future work. 

In the absence of experimental data, we use the calculation of \refcite{Lasenby:2021wsc} to assess the relevance of the detector geometry to the DM scattering rate, considering only the WSi detector layer and the immediately adjacent \ce{SiO2} layers. This calculation requires the dielectric function $\epsilon$ to be purely real outside the detector layer, meaning that these layers are dissipationless. We enforce this condition by explicitly taking the real part of $\epsilon$ outside the detector layer. This approximation is valuable to highlight a unique effect that takes place when the detector layer is much more strongly dissipative than the other layers: in this case, deposits in those other layers must be conducted to the detector layer before they can dissipate. This means that the detector is sensitive to deposits far from the detector layer, dramatically enhancing the effective volume of the system. This is also the reason for the integral in \cref{eq:thin-loss} to be extended over all space. Indeed, in the presence of dissipation in all space, this integral would diverge.

However, in our prototype, dissipation in the other layers is in fact non-negligible. Preliminary experimental results suggest that a deposit in another layer must be above the threshold by a factor of $\mathcal O(100)$ in order to reliably trigger the SNSPD, and understanding the effective available detector volume as a function of the deposited energy requires more detailed laboratory characterization. We thus show an additional conservative benchmark (dotted curves in \cref{fig:scat}) in which the dielectric function is allowed to be complex everywhere, but only deposits within the WSi detector layer are included, \textit{i.e.} the domain of the integral in \cref{eq:thin-loss} is restricted. In addition to the \ce{SiO2} layers, we include the ZEP520A layer, treating it as semi-infinite in extent. This simplistic estimate demonstrates that when $\epsilon$ is allowed to be complex everywhere, the scattering rate is enhanced even when deposits outside the detector layer are neglected. Ultimately, direct experimental characterization can eliminate uncertainty in our treatment of geometric effects for both scattering and absorption.

\bibliography{references}

\end{document}